\documentclass[reprint,showpacs,aps]{revtex4-1}
\usepackage{mathrsfs,amsmath,amssymb}
\usepackage{amsfonts}
\usepackage{graphicx}
\usepackage{float}
\usepackage{subfigure}
\usepackage{tikz}
\usetikzlibrary{arrows}
\usepackage{bm}
\begin{document}

\title{Extraordinary spin density and energy back-flow under interference}
\author{Zhen-Lai Wang, Dan-Dan Lian and Xiang-Song Chen}
\email{cxs@hust.edu.cn}
\affiliation{ School of Physics and MOE Key Laboratory of Fundamental Quantities Measurement, Huazhong University of Science and Technology, Wuhan 430074,  P. R. China}

\date{\today}

\begin{abstract}
A novel phenomenon was reported recently that the ``local optical spin density" based on Poynting vector might be counter-intuitively opposite to the integrated spin orientation while the one related to the canonical expression might not [Leader, Phys. Lett. B 779, 385-387 (2018)]. However, the ``local optical spin density" of the canonical expression can also be counter-intuitively opposite to the integrated spin orientation under the interference of plane waves, even if all of the plane waves possess the same polarization handedness. Moreover, the interference fields might acquire a transverse spin density (perpendicular to the propagation plane), which can have more well-controlled relations with the polarization. Meantime, in such a case its energy flux exhibits counter-intuitive back-flow and a circular motion (vortex) in the propagation plane locally, which  implies a transverse local orbital angular momentum density.

\end{abstract}

\maketitle

Light, or more generally an electromagnetic wave, can carry energy, momentum and angular momentum, and all these properties are well confirmed by light-matter interactions. Now it is well-known that a light beam can carry two distinct forms of angular momentum. One, called spin angular momentum, is related with the polarization of light; the other, called orbital angular momentum, is related with light's wave-front shape \cite{Allen92, Allen,Fran,Andr}. However, in the theoretical framework, it is not exactly clear how to handle the two different forms of angular momentum. One familiar expression to describle the optical angular momentum in classical electrodynamics \cite{Jack} is written as
 \begin{equation}\label{Js}
 {\bm J}=\int d^3 x [{\bm r}\times({\bm E}\times{\bm B})],
 \end{equation}
and therefore the total optical angular momentum density ${\bm j}_{\rm poy}={\bm r}\times({\bm E}\times{\bm B})= {\bm r}\times{\bm p}$ is expressed in an orbital structure with the Poynting vector ${\bm p}$ which is treated as the energy flux density of light:
 \begin{equation}\label{P}
 {\bm p}={\bm E}\times{\bm B}.
 \end{equation}
 
 Another expression of the optical angular momentum is commonly used in laser optics 
  \begin{equation}\label{Jc}
 {\bm J}=\int d^3 x {\bm E}\times{\bm A}+\int d^3 x  E^i({\bm r}\times{\bm \nabla}) A^i,
 \end{equation}
 which splits the total optical angular momentum into spin and orbital angular momentum, where ${\bm A}$ is the vector potential of  magnetic field. The expression given by Eq.(\ref{Jc}) can be derived from Noether's theorem in virtue of the space-rotation symmetry and is called canonical version of optical angular momentum. As Eq.(\ref{Jc}) manifestly shows, it is gauge-dependent and so not measurable in principle.
 A gauge-invariant version can be yielded by introducing the gauge-invariant part ${\bm A_{\perp}}$ of the vector potential ${\bm A}$ \cite{Coh,Van,Chen}: 
\begin{equation}\label{Jcg}
 {\bm J}=\int d^3 x {\bm E}\times{\bm A_{\perp}}+\int d^3 x  E^i({\bm r}\times{\bm \nabla}) A^i_{\perp},
 \end{equation}
Thus, Eq.(\ref{Jcg}) defines the total angular momentum density by ${\bm j}_{\rm can}={\bm s}_{\rm can}+{\bm l}_{\rm can}$ with spin angular momentum density
 \begin{equation}\label{jcs}
 {\bm s}_{\rm can}= {\bm E}\times{\bm A_{\perp}}
 \end{equation}
 and orbital angular momentum density 
 \begin{equation}\label{jcl}
{\bm l}_{\rm can}=E^i({\bm r}\times{\bm \nabla}) A^i_{\perp}={\bm r}\times{\bm p}_{\rm can}^{\rm o},
 \end{equation}
where the optical canonical momentum density reads
\begin{equation}\label{cpo}
{\bm p}_{\rm can}^o=E^i{\bm \nabla} A^i_{\perp}.
 \end{equation}
Here the subscript $i$ runs over the three spatial coordinate subscript.

For the free electromagnetic field, the integrated angular momentum of light in Eq.(\ref{Js}) and (\ref{Jcg}) are identical, but the angular momentum density ${\bm j}_{\rm poy}$ differs from ${\bm j}_{\rm can}$ by a surface term. Recently, basing on the measurability of the local spin and orbital angular momentum density, E. Leader \cite{Lea} proposed a measurement scheme for testing which one between ${\bm j}_{\rm poy}$ and ${\bm j}_{\rm can}$ is the effective expression of optical angular momentum density. Analysing a circularly polarised Laguerre-Gaussian beam in the paraxial approximation, he found that ${\bm j}_{\rm poy}$ and ${\bm j}_{\rm can}$ can yield different spin angular momentum densities. For ${\bm j}_{\rm poy}$ its $s_{\rm poy, z}$, the component of ``spin angular density" along the propagation direction, can display two different handedness, while for ${\bm j}_{\rm can}$ its $s_{\rm can, z}$ possesses a single handedness. 
 
In this work, we analyse the interference of three plane waves propagating in the same plane  with positive momentum in their common propagation direction ($z$-axis). The canonical spin density $s_{\rm can, z}$  can also display two different handednesses locally under interference of plane waves even if all of the plane waves possess the same polarization handedness along $z$-axis direction, in other words, $s_{\rm can, z}$ can locally reverse its handedness. Moreover, the interference field might possess a  transverse spin density (out-of-plane) which is perpendicular to the propagation plane and have more rich relations with the polarization.  Meantime, the Poynting vector can not only produce counter-intuitive back-flow, but also form a circular motion in the propagation plane, which implies a transverse orbital angular momentum(out-of-plane). 

 \begin{figure}
\includegraphics[width=0.4\textwidth]{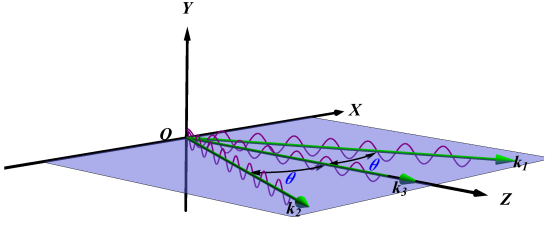}  
\caption{ Interference of three  plane waves all  propagating in the  $ x - z $ plane with equal amplitudes and 
       different wave vectors $ {\bm k}_1 $, ${\bm k}_2$ 
       and ${\bm k}_3$. The angle $\theta$ denotes the angle between  
     ${\bm k}_{1,2}$ and ${\bm k}_3$. }  
       \label{San0}
\end{figure}

Throughout this paper, we focus on monochromatic plane waves in free space.
Specifically,  we consider the interference of three plane waves propagating in
the  $ x - z $ plane. The electric and magnetic fields of the three waves are respectively 
\begin{eqnarray}\label{EB}
{\bm E}_a(\bm{r},t)& = &{\rm Re}[{\bm{\mathcal E}}_a(\bm{r})  e^{- {\rm i}\omega t}],\qquad\qquad(a=1,2,3)\\
{\bm{\mathcal E}}_a(\bm{r}) &=& E_0 e^{ {\rm i}{\phi _a}({\bm r})} {\bm \chi}_a,\qquad
{\bm{\mathcal B}}_a({\bm r}) = \frac{{\bm k}_a}{\omega }\times {\bm{\mathcal E}}_a({\bm r})\\
{\bm{\hat k}}_a\cdot{\bm e}_y&=&0,~{\bm \chi}_a=[{\bm e}_y + {\rm i}\sigma_a({\bm{\hat k}}_a\times {\bm e}_y)]/\sqrt{1+\sigma_a^2}
 \end{eqnarray} 
Here the phase factors ${\phi _a}(\bm {r}) = {\bm {k}_a} \cdot {\bm {r}}$  with 
$\bm {k}_a$  the wave vectors, ${\bm{\hat k}}_a={\bm k}_a/|{\bm k}_a|$ is the corresponding unit wave vectors, $E_0$ is the wave amplitude, ${{\bm e}_y}$ 
(after ${{\bm e}_z}$, ${{\bm e}_x}$) is the unit vector of $y$ ($x$, $z$) axis, and $\sigma\in [-1,1]$ determines the polarization state.
For simplicity, we set ${\bm k}_3$ in the $z$ axis and ${\bm k}_{1,2}$ to be distributed mirror-symmetrically about ${\bm k}_3$ in the $x-z$ plane, as depicted in FIG.\ref{San0}:
\begin{equation}
{\bm k}_{1,2} =  k(\cos {\theta } {\bm e}_z \pm \sin {\theta } {\bm e}_x),~~{\bm k}_3 =  k {\bm e}_z \nonumber,
 \end{equation}
 where $k = \omega/c=\omega$ (taking $c=1$) is the wave number, and ${\theta }$  denotes the angle between ${\bm k}_1$ (or ${\bm k}_2$)  and ${\bm k}_3$ ($\theta\in[0,\pi/2]$).
The resulting fields are 
$\bm {E} = \textstyle{\sum\limits _{a=1}^{3}}{{\bm {E}_a}(\bm {r},t)}$  and
$\bm {B} = \textstyle{\sum\limits _{a=1}^{3}}{{\bm {B}_a}(\bm {r},t)}$. 

For a monochromatic-wave case, we obtain the time-averaged Poynting vector, canonical momentum density and canonical spin density from Eq.(\ref{P})(\ref{jcs})(\ref{cpo}):
\begin{equation}
\overline{\bm p} =\overline{{\bm E} \times {\bm B}}= \frac{1}{2} {\rm Re}\left[ {\bm{\mathcal E}}^{*} \times {\bm{\mathcal B}}\right], \label{peb}
 \end{equation}
 \begin{equation}\label{cp}
\overline{\bm p} _{\rm can}^o=\overline{E^i{\bm \nabla} A^i_{\perp}}= \frac{1}{2\omega} {\rm Im}\left[ {\mathcal E}_i^{*} {\bm \nabla} {\mathcal E_i}\right], 
 \end{equation}
\begin{equation}\label{cs}
\overline{\bm s}_{\rm can}=2\overline{\bm s}_{\rm e}=\frac{1}{2 \omega}{\rm Im}\left[{\bm{\mathcal E^*}}\times{\bm{\mathcal E}}\right].
\end{equation}
Recently, the dual-symmetry between electric and magnetic fields has attracted considerable attention and the optical spin density was suggested  in the so-called  dual-symmetric form \cite{Cam,Bli13}:
\begin{equation}\label{ds}
\overline{\bm s}_{\rm dua}=\overline{\bm s}_{\rm e}+\overline{\bm s}_{\rm m}=\frac{1}{4 \omega}{\rm Im}\left[{\bm{\mathcal E^*}}\times{\bm{\mathcal E}}+{\bm{\mathcal B^*}}\times{\bm{\mathcal B}}\right].
\end{equation}
One can find that the canonical spin density in Eq.(\ref{cs}) is only determined by the electric field and that the dual-symmetric spin density in Eq.(\ref{ds}) coming from electric and magnetic field separately, so in this paper we introduce the so-called electric spin density ${\bm s}_{\rm e}$ and magnetic spin density ${\bm s}_{\rm m}$.  

Calculating from the  resulting electric field and magnetic field, we get the time-averaged Poynting vector, the electric and magnetic spin densities
\begin{eqnarray}\label{p}
\overline{\bm p} &=&\sum_{a,b=1}^{3}\frac{E_0^2}{2\sqrt{(1+\sigma_a^2)(1+\sigma_b^2)}}\Big[\cos\delta_{ab}(1+\sigma_a\sigma_b){\bm{\hat{k}}_a}\nonumber\\
&~&\qquad\qquad~~-\sin\delta_{ab}\sigma_a[({\bm{\hat{k}}_a}\times{\bm{\hat{k}}_b})\cdot\bm{e}_y]\bm{e}_y\Big]
\end{eqnarray}
\begin{eqnarray}\label{se}
\overline{\bm s}_{\rm e}&=&\sum_{a,b=1}^{3}\frac{E_0^2}{4\sqrt{(1+\sigma_a^2)(1+\sigma_b^2)}~\omega}\Big[\cos\delta_{ab} 2 \sigma_a{\bm{\hat k}}_a\nonumber\\
&~&\qquad~- \sigma_a\sigma_b\sin\delta_{ab}[({\bm{\hat k}}_a\times{\bm{\hat k}}_b)\cdot{\bm e}_y]\bm{e}_y\Big]
\end{eqnarray}
\begin{eqnarray}\label{sm}
\overline{\bm s}_{\rm m} &=&\sum_{a,b=1}^{3}\frac{E_0^2 }{4\sqrt{(1+\sigma_a^2)(1+\sigma_b^2)}~\omega}\Big[\cos\delta_{ab}2\sigma_a{\bm{\hat k}}_b\nonumber\\
&~&\qquad\qquad~-\sin\delta_{ab}[({\bm{\hat k}}_a\times{\bm{\hat k}}_b)\cdot{\bm e}_y]\bm{e}_y\Big]
\end{eqnarray}
Here $\delta_{ab}={\phi _a}-{\phi _b}$ are the corresponding phase differences, all of which vary in the  $ x - z $ plane. As we can see from Eqs.(\ref{p})-(\ref{sm}), the Poynting vector, the electric and magnetic spin densities have $y-$axis components while the resulting fields propagate entirely in the $x - z$ plane. More interestingly, the $y-$axis components of the electric and magnetic spin densities can be dual-asymmetric, one of which is dependent on the polarization and the other is independent on the polarization. Hence, the transverse electric and magnetic spin densities exhibit the strong electric-magnetic asymmetry. On the other hand, the transverse Poynting vector also depends upon the polarization state. On the topic of transverse optical spin density, there is a good review given by Bliokh and Nori \cite{Bli}. If  taking the resulting fields to take the electric-magnetic duality transformation, we will make the $y-$axis component of electric spin density (or the canonical spin density ${\bm s}_{\rm can}$) independent on the polarization but that of the magnetic spin density dependent on the polarization. In Ref \cite{Bek}, the transverse optical spin density yielded by interference of two plane waves might also show the electric-magnetic asymmetry, in which case the transverse electric and magnetic spin densities are actually exhibit different local distributions but are both independent on the helicity parameter $\sigma$. However, the strong electric-magnetic asymmetry presented here is due to the fact that the transverse electric and magnetic spin densities can have different relations with the polarization state. Usually the polarization-dependent spin angular momentum of light is the longitudinal component, but here we show a transverse (out-of-plane) polarization-dependent spin density.

\begin{figure}
 \centering
  \subfigure[]{\includegraphics[width=0.4\textwidth]{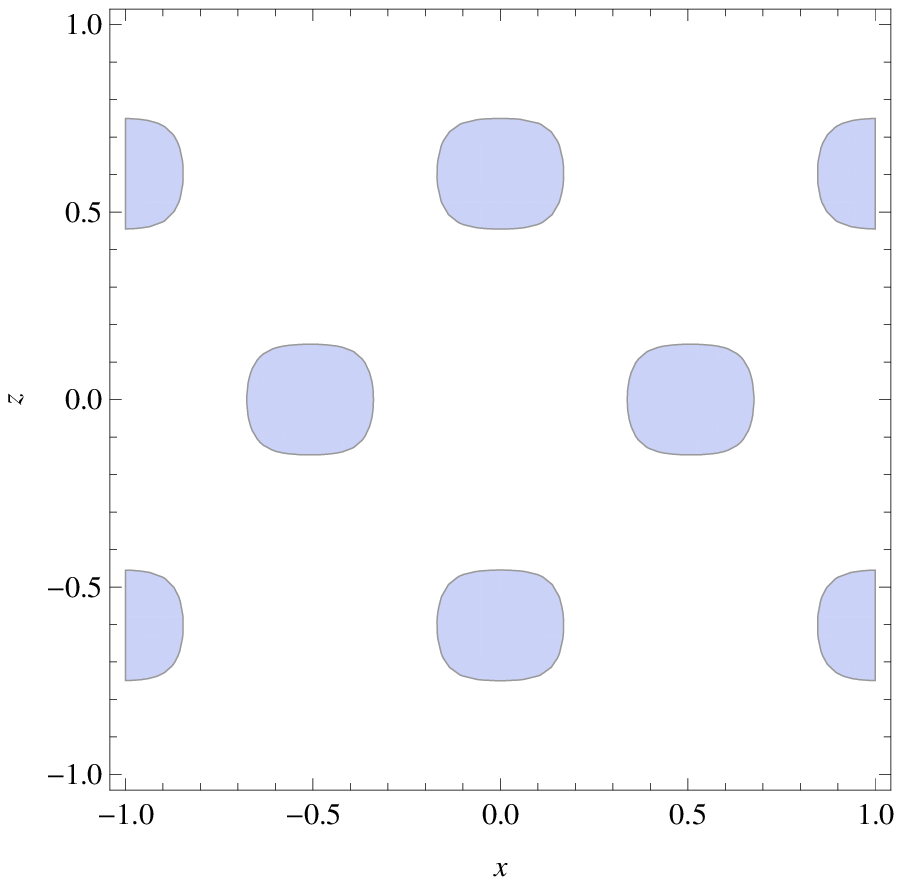}}\qquad
  \subfigure[]{\includegraphics[width=0.4\textwidth]{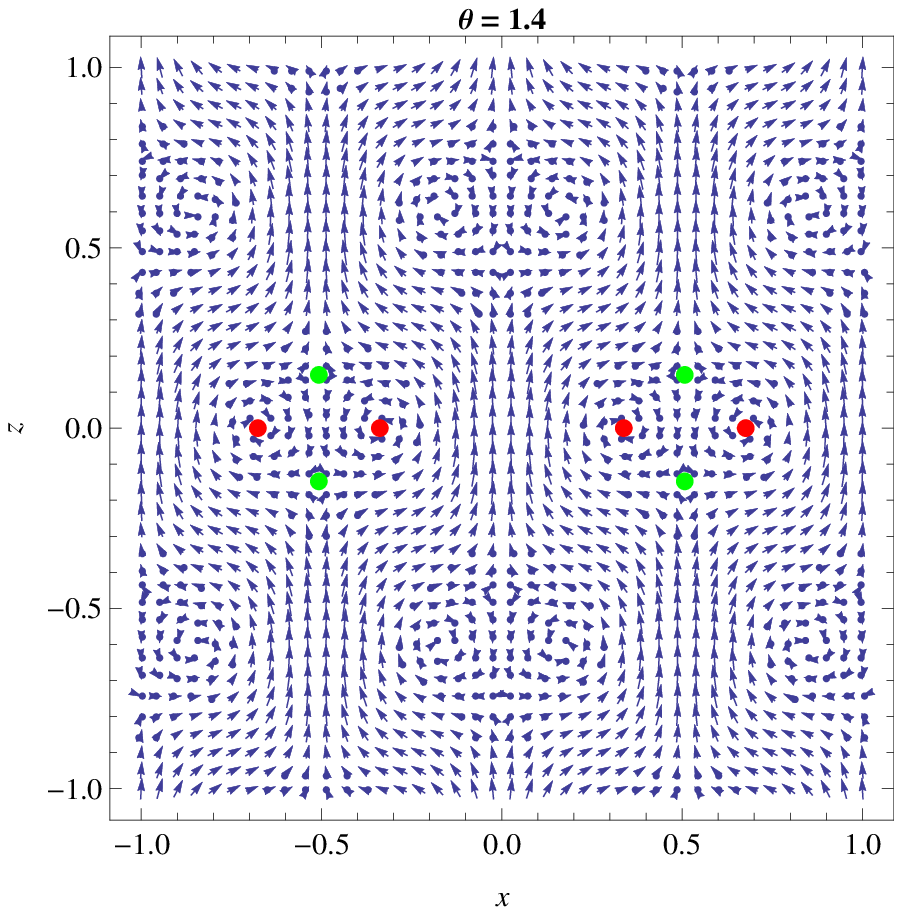}}\qquad
       \caption{ $ \overline{\bm p}$ and $\overline{\bm s}$ projected in the $ x - z $ plane when all $\sigma_a=1$. If all $\sigma_a=1$, the Poynting vector and spin density have the same structure; omitting their constant factors, here we plot the distribution of their common structure  (a)  Regions of negative ${\overline p}_z$  and ${\overline s}_z$ with $\theta=1.4$ and $ k=2\pi$ (b) Vector distribution of  $ ({\overline p}_x,{\overline p}_z)$ and $ ({\overline s}_x,{\overline s}_z)$ with $\theta=1.4$ and $ k=2\pi$. The red and green points are the singularity of $ ({\overline p}_x,{\overline p}_z)$ and $ ({\overline s}_x,{\overline s}_z)$. The circular motions take place around the red points.}
       \label{San2}
\end{figure}

More remarkably, from Eqs.(\ref{p})-(\ref{sm}) we can get that the Poynting vector and the spin density can reverse their direction along the main  propagating direction. For example, if all $\sigma_a=1$, the Poynting vector, the electric and magnetic spin densities have the same structure. For clarity, we plot the components of  $\overline{\bm p}$, $\overline{\bm s}_{\rm e}$ and $\overline{\bm s}_{\rm m}$ projected in the $x-z$ plane when all $\sigma_a=1$  in FIG.\ref{San2}. Seen from FIG.\ref{San2}, it is clear that ${\overline p}_z$  and ${\overline s}_z$ can be negative in some regions, telling that the Poynting vector could be opposite to the main propagation direction of the three incident waves and that the spin density can experience the transition of handedness locally. 

FIG.\ref{San2} shows vividly that the Poynting vector could  form a circle near the region of negative Poynting vector. In fact the circle is yeilded around a point where $\overline{\bm p} =0$, and the point is the so-called Poynting singularity \cite{Nov09,Bek11}. Naturally, the Poynting vector might change from a forward  to a reversed flow at the region of Poynting singularity. By the  expression of total angular momentum of a free electromagnetic field, Eq.(\ref{Js}), the circular motion of Poynting vector suggests the existence of local transverse (out-of-plane) angular momentum. Obviously, if all $\sigma_a=0$, i.e. all the three plane waves  are  linearly polarized in $y$ axis, the components of the Poynting vector $({\overline  p}_x,{\overline  p}_z)$ do keep the same structure as these components do when all $\sigma_a=1$. On the other hand, when all $\sigma_a=0$, the Poynting vector is identical to the optical canonical momentum. Thus, it is clear from Eq.(\ref{Jcg}) that the local transverse (out-of-plane) angular momentum does not contain spin ingredient but fully belongs to orbital angular momentum. Also, FIG.\ref{San2}(b) informs us that the a circular motion (vortex) always emerges in pairs with opposite rotation directions. They mutually cancel out each other and make the integrated transverse angular momentum vanish. Seen from the two figures in FIG.\ref{San2}, the reversed Poynting vector appears in the middle region of a pair of vortexes.

If all $\sigma_a=1$,  the electric and magnetic spin densities have the same structure as the Poynting vector, and they of cause form spin vortexes in the  $x-z$ plane. The reversed spin densities also appear in the middle region of a pair of vortexes, as depicted in FIG.\ref{San2}. All of these counter-intuitive phenomena are the direct result of wave interference effect.

To observe the backflow and circulation of the Poynting vector (or momentum density) and the extraordinary spin density, we need a detection scheme sensitive to local energy flow (or momentum density) and spin density. One possible experiment could involve optical forces and torque on small particles. Usually, an absorptive and nonmagnetic particle (Rayleigh particle) with dimensions much smaller than the wavelength of the scattered light is well described by an electric-dipole \cite{Bek,Ash,Cha,Alb,Can,Bek13,Bli14,Luk}. From this effective model, one can get the time averaged optical force $\bm {f}$ and torque $\bm {\tau}$ on the particle:
\begin{equation}
\bm {f} =  \frac{\mathop{\rm Re}\nolimits{(\alpha)} } {2}~\bm {\nabla } w^e+\frac{\omega}{2}  {\mathop{\rm Im}\nolimits{(\alpha)}}~\overline{\bm p} _{\rm can}^o,~~
 \bm {\tau} ={\mathop{\rm Im}\nolimits{(\alpha)}}\overline{\bm s} _{\rm e}, 
\end{equation}
where $w^e=\overline{{\bm E}^2}/2$, the first term in the optical force  expression stands for the gradient force and the second one the scattering force.  Here the polarizability $\alpha $ of the particle depends on the nature of the particle. Therefore, the extraordinary spin density and Poynting vector (is equal to optical canonical momentum $\overline{\bm p} _{\rm can}^o$ if all $\sigma_a=0$.) can be directly measured by the local interaction between light and the test-particle.

In conclusion, we have shown the ``local optical spin density" of the canonical expression also  might be counter-intuitively opposite to the integrated spin orientation under interference of plane waves. Further analysis shows that the interference fields can produce an extraordinary Poynting vector, which can not only be negative (the energy flux is opposite to the propagation direction), but also form a circular motion (vortex) and result in local transverse (out-of-plane) orbital angular momentum. Furthermore, this simple optical system also acquires a transverse spin density which shows strong electric-magnetic asymmetry; for the  transverse electric and magnetic spin densities, one of them is polarization-dependent and the other is polarization-independent.  All of these counter-intuitive findings are the result of wave interference effect, which should have useful implications in fundamental studies and applications of light.

This research is supported by  the China NSF under Grants No. 11535005  and No. 11275077.

\end{document}